# Home Electricity Data Generator (HEDGE): An open-access tool for the generation of electric vehicle, residential demand, and PV generation profiles

*Flora Charbonnier[1]\*, Thomas Morstyn[2], Malcolm McCulloch[1]*

## Keywords

Datasets, Data-driven methods, Open access, Demand-side response, Distributed energy resources, Buildings, Smart grid, Residential

## Abstract

In this paper, we present the *Home Electricity Data Generator (HEDGE)*, an open-access tool for the random generation of realistic residential energy data. HEDGE generates realistic daily profiles of residential PV generation, household electric loads, and electric vehicle consumption and at-home availability, based on real-life UK datasets. The lack of usable data is a major hurdle for research on residential distributed energy resources characterisation and coordination, especially when using data-driven methods such as machine learning-based forecasting and reinforcement learning-based control. A key issue is that while large data banks are available, they are not in a usable format, and numerous subsequent days of data for a given single home are unavailable. We fill these gaps with the open-access HEDGE tool which generates data sequences of energy data for several days in a way that is consistent for single homes, both in terms of profile magnitude and behavioural clusters. From raw datasets, pre-processing steps are conducted, including filling in incomplete data sequences and clustering profiles into behaviour clusters. Generative adversarial networks (GANs) are then trained to generate realistic synthetic data representative of each behaviour groups consistent with real-life behavioural and physical patterns.

[1] Department of Engineering Science, University of Oxford, UK.
\*Corresponding author: flora.charbonnier@eng.ox.ac.uk
[2] School of Engineering, University of Edinburgh, UK.

Graphical abstract

### Input data

| 2020-10-08T23:01:00Z | Car | 3.2 miles | ... |
| 2020-12-02T10:33:00Z | Bus | 5 miles | ... |
| 2020-12-03T14:12:00Z | Bike | 0.9 miles | ... |
| ... | | | |
| ... | | | |

List of punctual trips

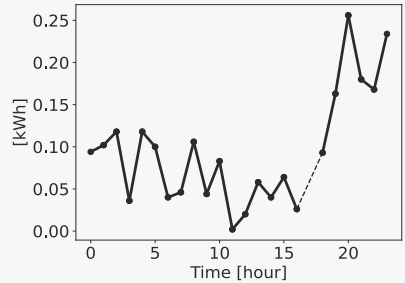

Incomplete generation and loads profiles

**Section 2: Data processing**

### Intermediate dataset

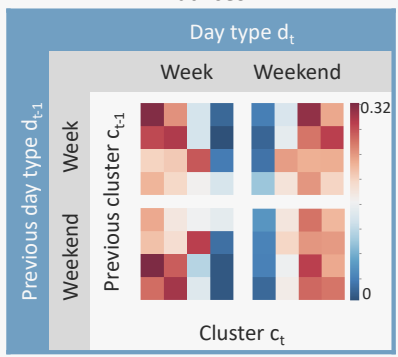

(A) Behaviour cluster transition matrices

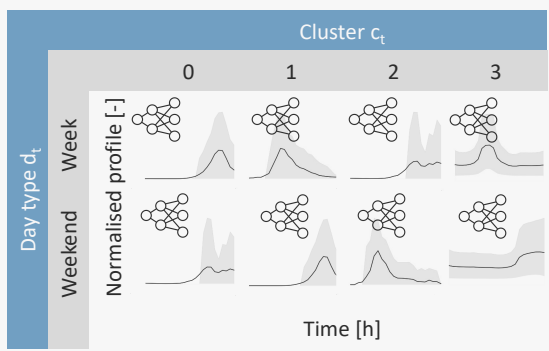

(B) Normalised profiles generators

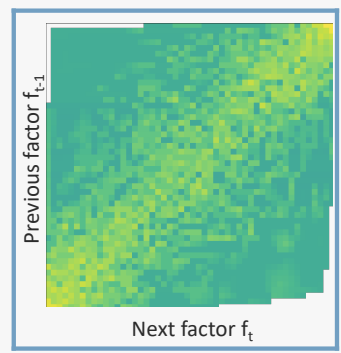

(C) Scaling factors transition matrices

$c_{t-1}, f_{t-1}, d_{t-1}, d_t$ — **Section 3: HEDGE tool** — Profiles drawn from $(d_t, c_t)$ and scaled by $f_t$

### Daily profiles

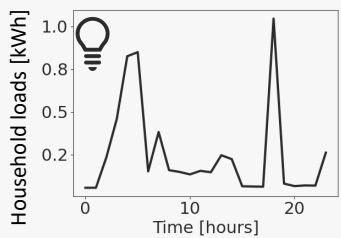
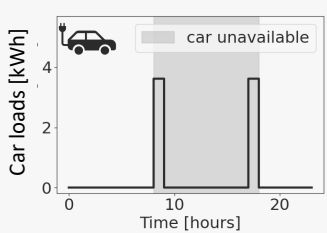
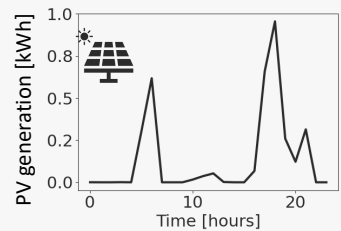

# 1. Objectives and Motivation

The Home Energy Data Generator (HEDGE) tool tackles the challenge of how to generate home energy consumption and generation data for use in data-driven algorithms. This open-access tool[3] can generate realistic photovoltaic (PV) generation, household loads, and electric vehicle (EV) consumption and at-home availability profiles.

The characterisation and simulation of residential energy resources is of increasing interest given their potential for demand-side response [1]. Renewable energy could supply 70% to 85% of electricity globally by 2050 in 1.5°C-compatible pathways[4][2], with corresponding needs for storage and demand-side response [1]. Demand-side response is also critical for the electrification of residential heating, cooling and transport, which, without coordination, could cause a significant increase in peak electricity demand with adverse consequences for low-voltage distribution networks [3]. Residential consumers could play an essential role in providing demand-side response [4], given as much as 53% of household demand could be flexible in the future [5].

Data-driven methods are of particular interest in the field of residential energy forecasting and control for three main reasons. Firstly, there is high uncertainty at the local level, due to the small scale of residential of electricity consumption and generation, and their behavioural and weather dependencies. Secondly, there are limitations to personal data sharing, particularly in real-time. This is due to both the limited availability of communication and computation infrastructure at the scale of an individual homes and to the privacy requirements of the residential sector. Thirdly, centralised optimisation methods have limited scalability. Therefore, data-driven analysis and control of the residential energy sector are of increasing interest [6], [7]. Figure 1 thus shows that the number of publications in the field has increased exponentially since 2000 (30% average yearly increase).

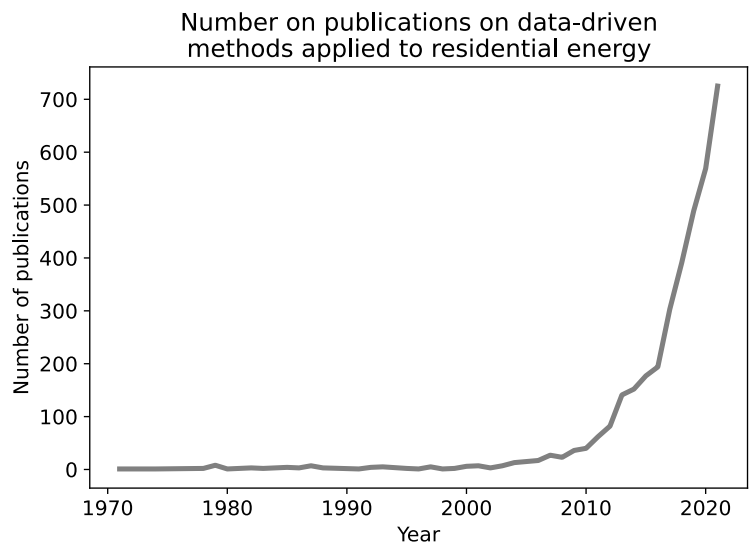

*Figure 1: Number of publications on data-driven methods for residential energy[5]*

A major hurdle for the development and implementation of such data-driven methods is the availability of large datasets on EV consumption and at-home availability, PV generation, and household consumption for training and testing data. The quality of data determines the results of data driven methods such as machine learning (ML) predictions or reinforcement learning (RL) policies [8], and should be as much of a focus as algorithm selection. While large amounts of residential energy data are indeed collected, training directly on available data is often unsatisfactory given:

---

[3] https://github.com/floracharbo/hedge
[4] interquartile range with no or limited overshoot (high confidence)
[5] Scopus key word search selecting for publications whose title and abstract include at least one mention of each the residential sector (home or residential), of data-driven methods (data-driven, learning, big data, or forecasting), the energy sector (energy, electricity, power, voltage, renewable) and local energy appliances (car, PV, solar, loads, smart, IoT, storage, battery, heating, HVAC, generation, electric vehicle, EV, appliance, demand response, demand-side response, peer-to-peer, consumer, or fridge)

- The privacy and costs constraints of data collection, or cost of access to datasets that are not freely available without a licence or privileged access. Obtaining energy data frequently poses a significant challenge for the development of energy communities [9]. This can result in substantial time and financial resources being expended. Generally, open-access databases offer rather restricted access to comprehensive energy consumption and production profiles, as the establishment of open-access data initiatives is fraught with numerous legal and occasionally ethical obstacles and inquiries. Companies may be wary about sharing their energy data outside of their business [9].
- The limited number of years of data collection available (e.g., for electric cars, for which we only have smart trial data from early adopters), or the limited number of subsequent days of data available for a given household, which hinders consistent simulation of a home for more extended. For example, the National Travel Survey offers at most a week of travel data for a given household [10].
- The labour-intensiveness of pre-processing of data, with efforts repeated across individual projects, as datasets are often not in a usable format, or not self-consistent across different days. Data quality has thus been identified as a challenge for the adoption of AI in the smart energy industry [11]. A major hurdle identified by energy community initiators is thus that of data formatting standards and the quality of the acquired data [9].

While agent-based modelling approaches have been previously adopted to model residential data such as EV patterns [12], training data should reflect real-life resource intermittency and behaviour variability to minimise training losses in a robust way without over-fitting. Purely synthetic data often lacks these characteristics. Moreover, bottom-up models such as CREST [14] rely on assumptions on dwelling activities and thermal-electrical demand modes for generating data.

Therefore, novel methods are required to meet the needs of both large-scale datasets and the inclusion of real-life patterns. A standard residential energy data generation tool that could interface with a local energy system benchmarking environment to generate continuous daily energy data for several days in a consistent manner, both in terms of profile magnitude and behavioural clusters, would greatly benefit the research community.

We bridge this gap by proposing a new tool which generates EV-, PV- and household demand-related data semi-randomly based on large-scale real-life datasets, while preserving profile magnitude and behavioural consistency over time. Compared to [13], which first proposed the use of generative adversarial networks to generate smart grid-related data, we further provide a data generator for UK data, integrate this tool directly into a MARL benchmarking framework, and include EV data generation. While this model uses UK data, the model could be adapted to use similar data from other countries, so long as banks of data of household consumption, PV generation, and travel patterns are available.

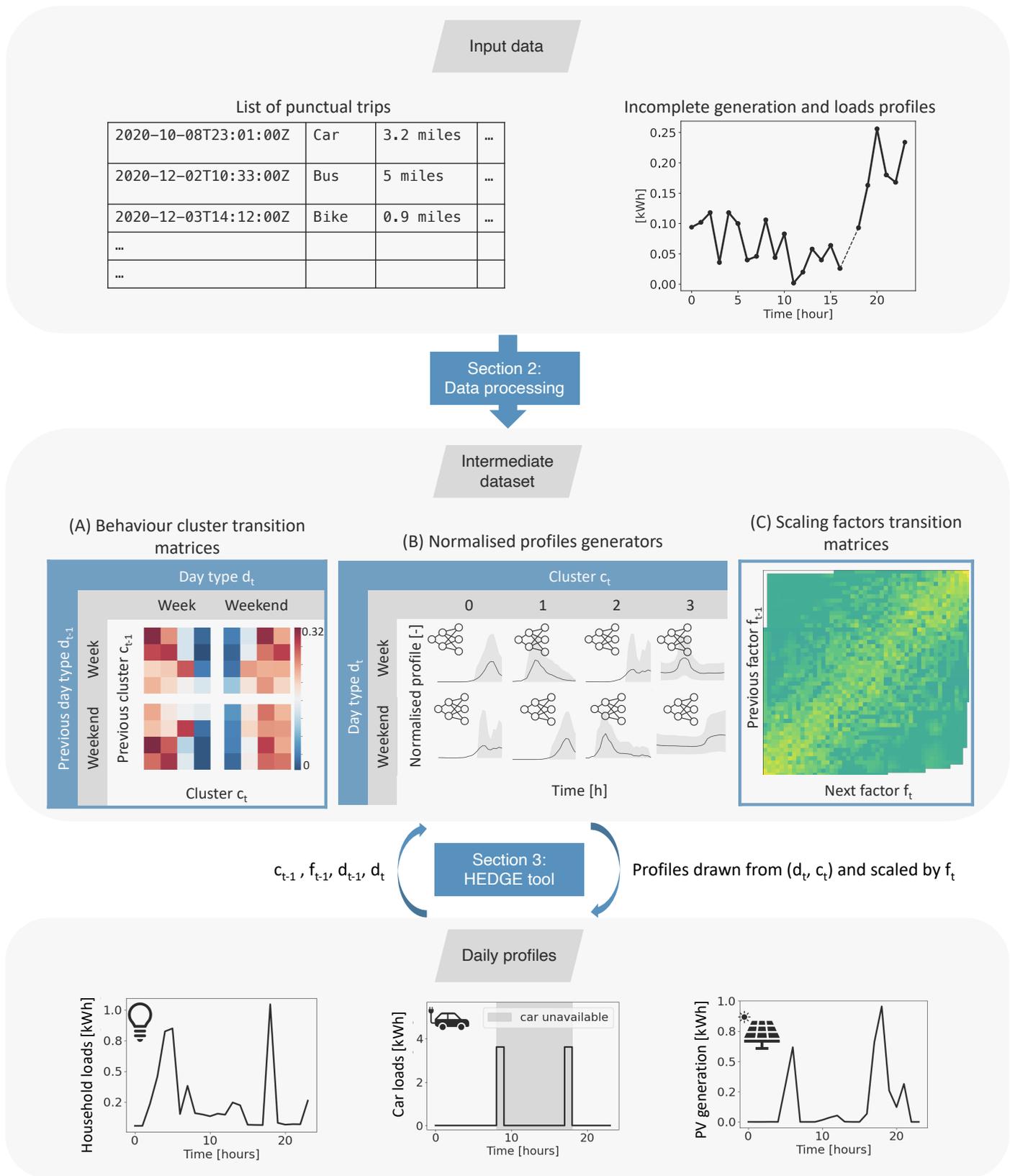

*Figure 2: Workflow from raw input data to the generation of random realistic household energy profiles. The two main steps "Data processing" and "HEDGE tool" use correspond to the sections 2 and 3 of this MethodX document.*

The rest of this MethodX paper is structured as illustrated in Figure 2. In Section 2, we present the data pre-processing steps to obtain the intermediate data used by the HEGE tool. In Section 3, we then present the mechanism used by HEDGE to generate data profiles. Finally, we list some limitations of this method in Section 4.

## 2. Data preparation

The data preparation steps are listed in Table 1 and detailed in the subsections below.

*Table 1: Data preparation steps*

| Step | Solar generation | Household loads | Electric vehicles |
|---|---|---|---|
| 1. Import data sources | Customer-led network revolution (CLNR) dataset TC1a [14] | CLNR dataset TC5 [15] | UK National Travel Survey [10] |
| 2. Data selection and filtering | Only residential data is used, and for valid date ranges. | | Only residential car journeys are selected. |
| 3. Conversion to relevant daily profiles | Convert to resolution specified (which has to be greater than or equal to 1 minute) | Get resolution specified (greater than or equal to 30 mins) | Convert list of trips to the distance travelled per time interval at the resolution specified. Infer the type of trip (motorway, urban, rural) from the location and distance. Then convert the distance travelled to electricity consumption based on the trip type. Infer at-home availability of the car based on trip times, origins and destinations. |
| 4. Missing data interpolation | Linearly fill in single missing time steps or discard the day of data. | | |
| 5. Normalisation | Normalise daily profiles by the sum of the electricity consumption/generation over one day. Record the scaling factors. | | |
| 6. Behaviour grouping | No clustering – group by month. | For each day type (weekday and weekend day), obtain 4 clusters using K-means. | For each day type (weekday and weekend day), obtain 3 clusters using K-means, as well as one for no-travel days. |
| 7. Profile generation | Train generative adversarial networks (GANs) to generate realistic profiles for each behaviour group. | | |
| 8. Scaling factor transition characterisation | Using 50 discrete time intervals for each day type transition, obtain the discrete transition probabilities between subsequent days. | | |
| 9. Behaviour cluster transition characterisation | No clustering. | Compute the transition probabilities for each cluster type and day type transition based on the real datasets. | |

## 2.1. Data import

Load and PV generation profiles are obtained from the Customer-Led Network Revolution (CLNR), a UK-based smart grid demonstration project [14], [15] which collected data from 13,000 customers between 2011 and 2014. PV sources have nominal capacities between 1.35 and 2.02 kWp.

We use mobility data from the National Travel Survey (NTS) [10] from 105,912 Great Britain households between 2002 and 2020. The NTS surveys the general population's travel patterns and does not focus on EVs – we have selected this dataset rather than an EV trial data, as this offers a less biased view into the general population's travel patterns thanks to both the larger volume of data available, and because the self-selected EV early trial participants may not be representative of patterns once EVs become widely adopted. We assumed that internal combustion engine (ICE) car travel patterns can be substituted for those of EVs, within battery constraints [16].

To overcome memory issues as well as limit computational time, the datasets are broken down into $n$ segments, without interrupting data for single homes. Data size reduction steps such as data filtering and granularity adjustments are conducted first before merging the different streams.

A limitation of these datasets is that behaviour and load profiles may have evolved since the date of collection. For example, the use of incandescent rather than LED lights was more common historically, and work patterns have evolved. Moreover, the datasets were collected in the UK, and may not be representative of other countries. However, the methodology proposed could be used with other datasets for different contexts.

### 2.2. Data selection and filtering

Firstly, the measurements of interests are selected. In the case of the NTS data, only household car trips are conserved, and only homes that can be classified as urban and rural are used. This is because the household type is needed to infer driving type and convert trips into electricity use at a later stage. Moreover, we remove trips above maximum user-defined hourly and daily energy demand, which would not be feasible with an electric car.

Then, the start and end times for data validity for each home are enforced and data beyond valid ranges discarded. Data validity ranges are characterised by the start of valid time, the end of valid time, and the duration of valid time. If one of these is missing, it can be inferred. If two or more of these pieces of information are missing, the validity of data cannot be confirmed, and it is discarded.

### 2.3. Conversion to relevant daily profiles

Sequences of subsequent data points for single homes are converted to the required resolution (e.g., hourly), and split into individual days.

In the case of CLNR data, this time granularity must be lower than that of the original data, e.g., one minute for PV generation and 30 minutes for household loads. Incomplete days with more than one consecutive data point missing are discarded.

In the case of the NTS travel data, lists of trips are converted to daily profiles of distance travelled. The at home-availability of the vehicles is then inferred from the recorded journeys' origin and destination. Equivalent EV energy consumption profiles are obtained using representative consumption factors from a tank-to-wheel model proposed in [16], dependent on travel speed and type (rural, urban, motorway). Motorway travel is assumed for trips larger than 10 miles.

### 2.4. Missing data interpolation

To fill in single missing data points, we test the following options:

1. Linearly interpolate between time steps before and after
2. Replace with the datapoint at the same time the day before or after (whichever has the lowest sum of squares of differences between the previous and subsequent point on the current day)
3. Replace with the datapoint at the same time one or two days before or after (whichever has the lowest sum of squares of differences between the previous and subsequent point on the current day)
4. Replace with the datapoint at the same time one day or week before or after (whichever has the lowest sum of squares of differences between the previous and subsequent point on the current day)

As shown in Figure 3, linearly interpolating results in the lowest average and $99^{th}$ percentile. We therefore use this method to fill in missing data.

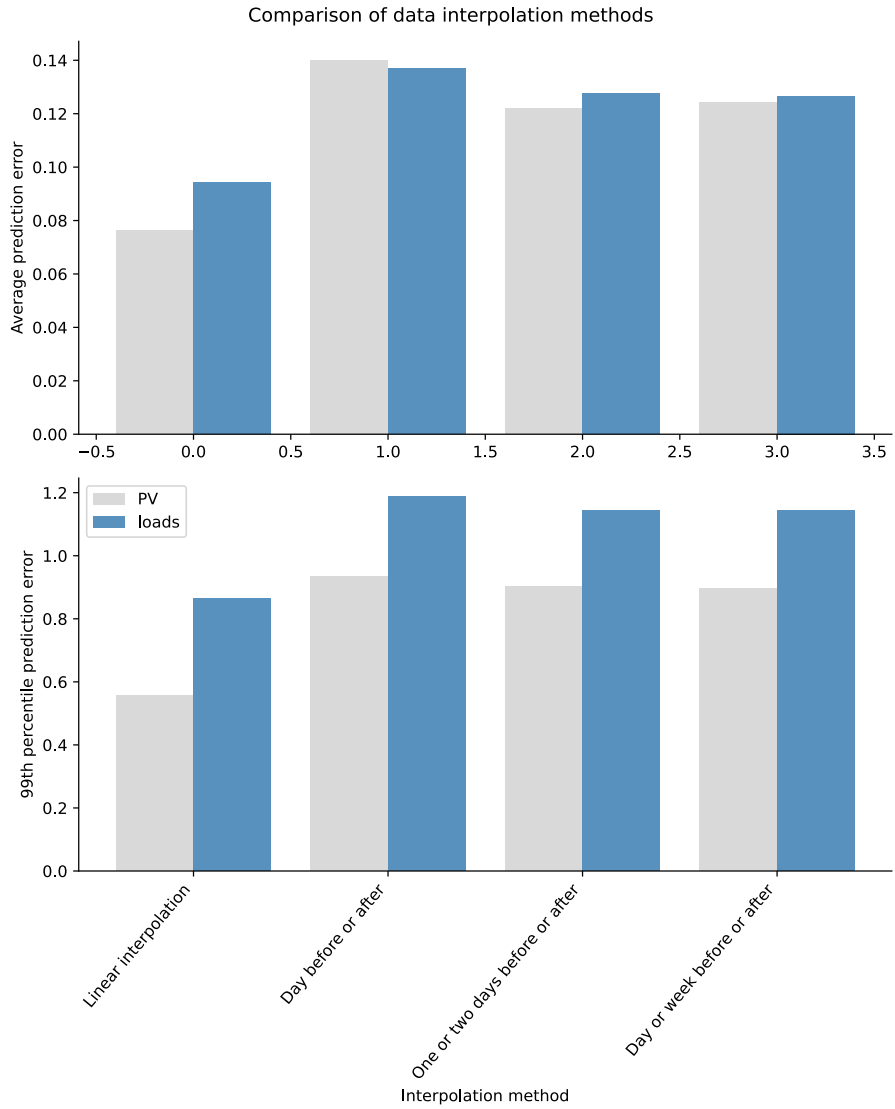

*Figure 3: Comparison of interpolation methods*

## 2.5. Normalisation

Normalisation is performed ahead of profile clustering and GAN training. Each daily profiles for energy generation and consumption are normalised such that $\sum_{t=0}^{24} x(t) = 1$, and the corresponding scaling factors are recorded.

These profiles can then be scaled up consistently to match the expected total energy generation/consumption over a day for a given household by the generation tool, as further described in Sections 2.8 and 3.

## 2.6. Behaviour clustering

For behaviour-dependent profiles, specifically household loads and EV patterns, the normalised profiles are grouped into clusters based on behavioural patterns for both weekday and weekend days. This clustering facilitates the creation of a repository of normalised profiles for each cluster group. This collection of profiles can subsequently serve as the foundation for training GANs to generate profiles representative of each cluster. When using HEDGE, different homes will have different likelihood of belonging to each behaviour group, and profiles can be generated accordingly to maintain consistency.

We use K-means, minimising the within-cluster sum-of-squares [17] in four clusters for both weekday and weekend data (with one for no travel). The features used for load profiles clustering are normalised peak magnitude and time, and normalised values over critical time windows[6], and those for travel are normalised values between 6 am and 10 pm. PV profiles were grouped per month. The user can define the number of clusters as an input.

As an example, the weekday behaviour clusters for household load and EV consumption are illustrated in Figure 4 and Figure 5.

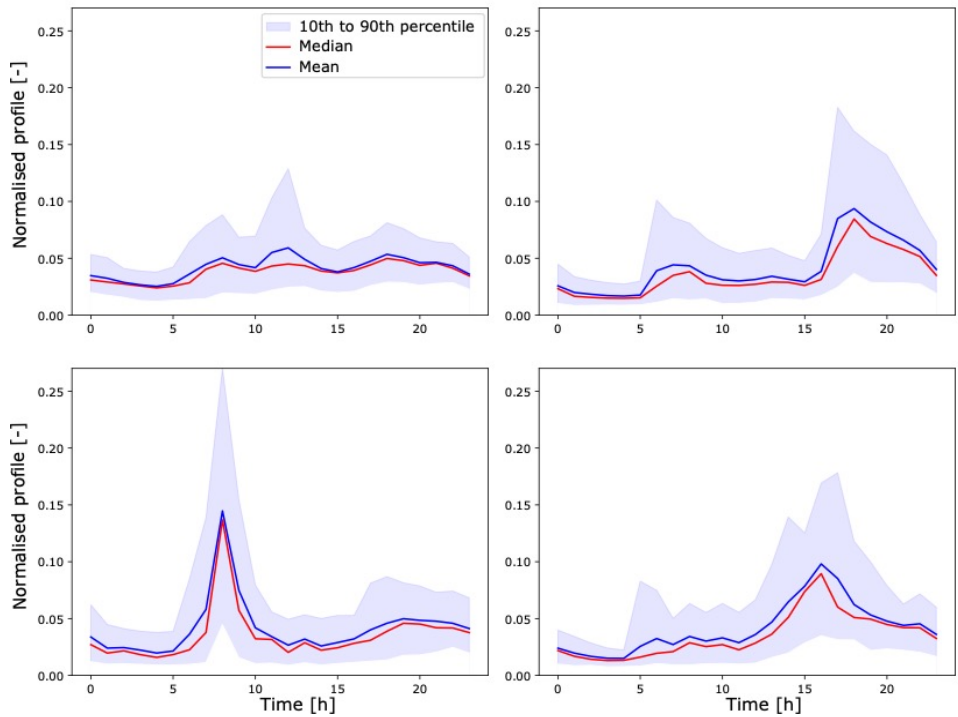

*Figure 4: Four behaviour clusters for weekday household electricity consumption normalised profiles*

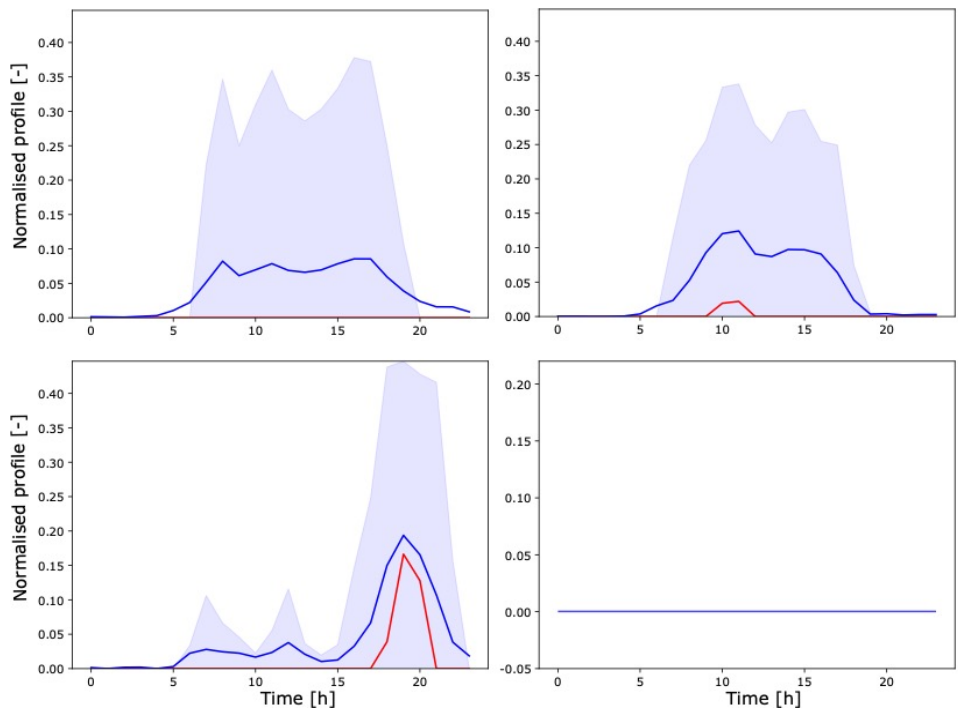

*Figure 5: Four behaviour clusters for weekday EV consumption normalised profiles. The fourth cluster corresponds to days with no travel.*

## 2.7. Profiles generation

Neural networks are then trained to generate populations of realistic normalised profiles corresponding to each behaviour cluster and day type. Pre-training neural network weights means that researchers and practitioners do not need to download large databases (here, the raw databases that had to be downloaded were of size 40.12 GB) and run time- and computational resource-hungry data preparation and training steps. They only need to download the pre-trained weights (files of size 125 kB) and perform a feed-forward to generate realistic training and testing data using HEDGE.

---

[6] 0-7 am, 7-11 am, 11 am-2 pm, 2-5 pm, 5-9 pm, 9-12 pm

As illustrated in Figure 6, GANs [18] consist of two simultaneously trained models. The generative model $\mathcal{G}$ takes as input a random noise vector $z$ and produces fake data $x_{synthetic} = \mathcal{G}(z)$, aiming to fool the discriminator into thinking they are from the original dataset $x_{real}$. The discriminator model $\mathcal{D}$ takes as input data $x$ and produces a probability score $\mathcal{D}(x) \in [0,1]$ that indicates the likelihood that $x$ is real data.

Each network aims to minimise the following losses during training:

- The discriminator $\mathcal{D}$ aims to maximise the probability of correctly discriminating between the real real data and the fake data generated by the generator network $\mathcal{G}$, by minimising the binary cross-entropy between the real (1) and fake (0) labels and the probabilities assigned by the discriminator:

$$\ell_{\mathcal{D}} = -\mathbb{E}_{x_{\text{real}}}[\log \mathcal{D}(x_{real})] - \mathbb{E}_z[\log(1 - \mathcal{D}(\mathcal{G}(z))])$$

- The generator loss is calculated from the discriminator's classification – It gets rewarded if it successfully fools the discriminator and gets penalised otherwise. The loss function aims to minimise the binary cross-entropy between the fake labels and the probabilities assigned to the fake generated data by the discriminator:

$$\ell_{\mathcal{G}} = -\mathbb{E}_z[\log(\mathcal{D}(\mathcal{G}(z)))]$$

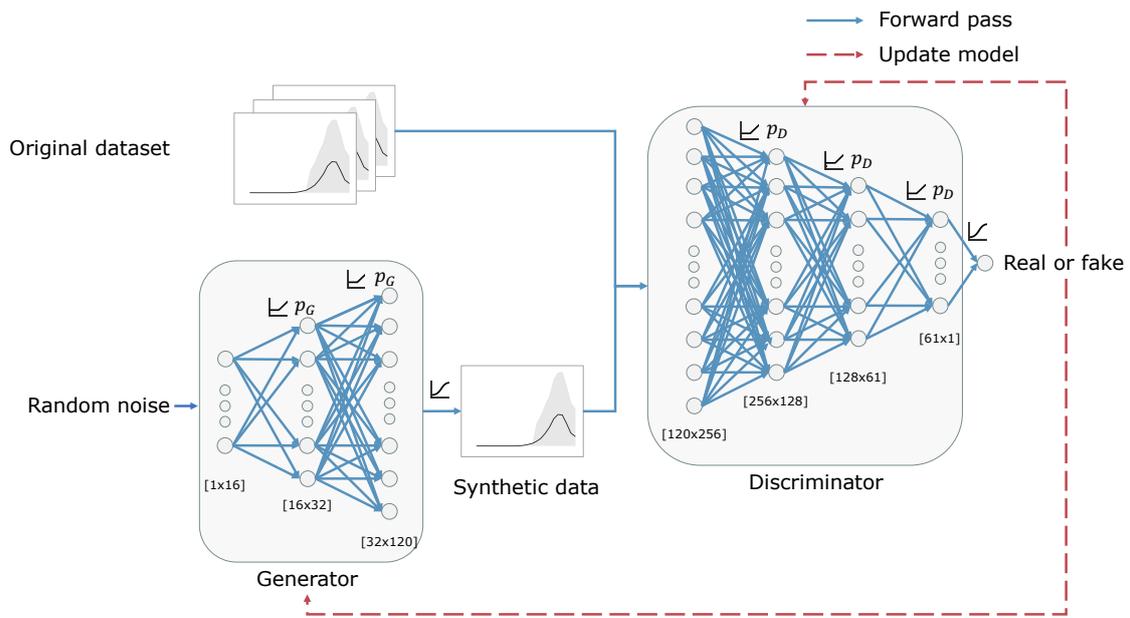

*Figure 6: Generative adversarial networks architecture for a given behaviour cluster and day type.*

To generate realistic populations of synthetic profiles, we use the following additional parameters and algorithm configurations:

- We exponentially decay the learning rate to avoid oscillation and to obtain faster convergence [19], so that the learning rate at each epoch is:

$$\alpha_{epoch} = \alpha_0 \left(\frac{\alpha_{\text{end}}}{\alpha_0}\right)^{\frac{epoch}{n_{\text{epochs}}}}$$

- We enforce the positivity of the generator's output by using the sigmoid activation function [20] on the last layer on the generator network:

$$\sigma(x) = \frac{1}{1 + e^{-x}}$$

- We employ dropout layers [21] within the neural network architectures to improve the performance of the models. This prevents overfitting to the training data by randomly dropping out (setting to zero) some of the outputs of the neurons during training, with probability $p_G$ for the generator and $p_D$ for the discriminator, effectively removing them from the network for that iteration. By doing this, the network becomes less sensitive to the specific weights of individual neurons and is forced to learn more robust features that are shared across multiple neurons.

Moreover, we further propose the following:

- We generate a population of profiles $i \in \{1, ..., n\}$ at each forward, pass, rather than one profile. This is to ensure that the GAN generates variability within one population that is realistic, rather than converging towards one realistic profile.
- We add an exponentially decaying noise to the exploration, to improve the efficiency and effectiveness of learning by encouraging exploration, avoiding overfitting oscillation, and obtain faster convergence. The decay helps balance the exploration and exploitation trade-off over time. The noise at each epoch is thus:

$$\epsilon_{epoch} = \epsilon_0 \left(\frac{\epsilon_{end}}{\epsilon_0}\right)^{\frac{epoch}{n_{epochs}}}$$

- We add a penalty to the generator's loss if the sum of the generated normalised profiles diverges from 1:

$$\ell_1 = W_1 \left(\frac{\sum_i \sum_t x_i^t}{n} - 1\right)^2$$

- We add a penalty to the generator's loss if the 10$^{th}$, 25$^{th}$, 50$^{th}$, 75$^{th}$ and 90$^{th}$ percentiles and the mean over the whole generated population for each time step $t$ varies from the original dataset for each time step:

$$\ell_2 = W_2 \sum_{k \in \{10^{th}, 25^{th}, 50^{th}, 75^{th}, 90^{th}, mean\}} \sum_t (x_k^t - x_{real}^t)^2$$

And $\ell'_G = \ell_G + \ell_1 + \ell_2$

Training parameter values are tabulated in Table 2.

*Table 2: Generative adversarial network training parameters.*

| | | | |
|---|---|---|---|
| Initial noise $\epsilon_0$ | 1 | Batch size $m$ | 100 |
| End noise $\epsilon_{end}$ | 1e-4 | Number of epochs $n_{epochs}$ | 200 |
| Initial learning rate $\alpha_0$ | 1e-2 | Number of profiles in generated population $n$ | 50 |
| End learning rate $\alpha_{end}$ | 1e-3 | Discriminator dropout probability $p_D$ | 0.3 |
| Normalised profiles loss weight $W_1$ | 0.1 | Generator dropout probability $p_G$ | 0.15 |
| Percentile distance loss weight $W_2$ | 100 | | |

### 2.8. Assessment of Generative Adversarial Networks

Assessing the performance of GANs can be challenging, especially for GANs generating time-series data, which is a more nascent field of study relative to the computer vision domain. A combination of both qualitative and quantitative assessments is recommended [22].

Firstly, we therefore perform a qualitative visual assessment of the profiles generated by the GAN. An example of a generated population of 50 household load profiles throughout the training is presented in Figure 7. While the profiles generated before the training starts do not match the target distribution, the population of profiles that is generated at the end of the training visually matches the target population in terms both of mean and in terms of the distribution and variability of the population of profiles throughout the day. This shows that the generated profiles are diverse enough, as samples are distributed to cover the real data.

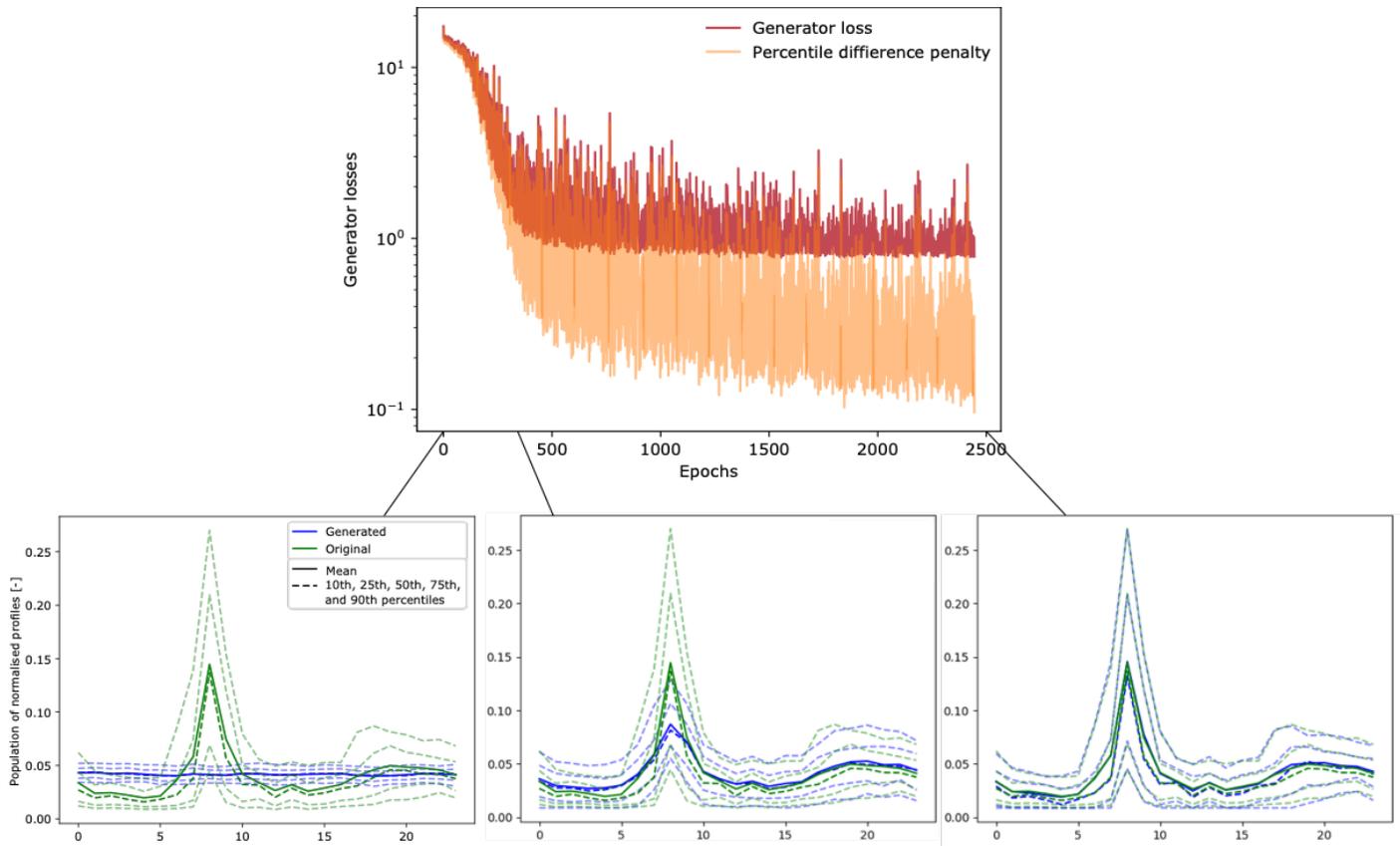

*Figure 7: Example of generated populations of 50 household load normalised profiles against the distribution of the original dataset throughout the GAN training.*

Secondly, we perform a quantitative evaluation, by adopting the "Train on Synthetic, Test on Real" (TSTR) framework proposed in [23] to evaluate the output of a GAN. This framework tests the usefulness of the GANs, by assessing the extent to which the generated data maintains the predictive attributes of the original. The testing sequence is as follows:

1. Split the real dataset into a training (80% of the data) and a testing (20%) dataset.
2. Train the GANs using the training dataset.
3. Generate synthetic data with the GANs.
4. Train a model using the synthetic data – Here, we train a classifier which aims at predicting which cluster a population of data profiles belongs to.
5. Test the classifier model using the held-out testing data. By determining the classifier's quality, this evaluation method, in turn, thus aims at assessing the quality of the generated data in being used for real applications.

Similar to the TSTR method, we also consider the reverse case, called "Train on Real, Test on Synthetic" (TRTS). Steps 1, 2 and 3 are identical, and steps 4 and 5 are interchanged as:

4. Train the classifier using the held-out testing data.
5. Test the classifier model using the synthetic data.

The performance of the classifiers in the TSTR and TRTS experiments presented in Figure 8 shows that the synthetic data generated by the trained GANs is useful for subsequent applications.

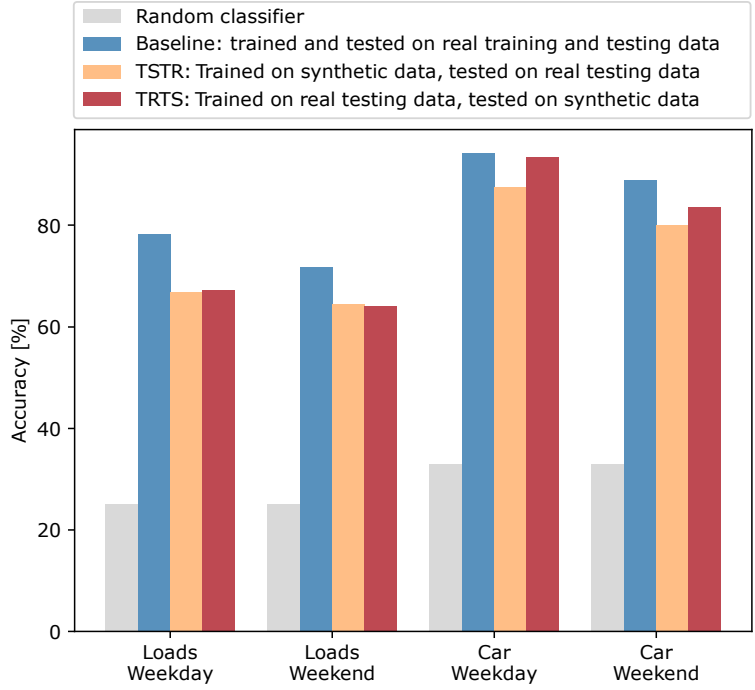

*Figure 8: "Train on Synthetic, Test on Real" and "Train on Real, Test on Synthetic" accuracy scores using the trained GANs relative to random and baseline classifiers. Average accuracy over 10 repetitions.*

## 2.9. Scaling factor transition characterisation

The unit-less normalised profiles generated by the trained GAN networks must then be scaled by a scaling factor consistent with a given home to produce profiles in energy units.

We use transition matrices to model the probability of transitioning from one scaling factor $f_t$ to the next one $f_{t+1}$ in subsequent days. Using these matrices allows the data generator to scale subsequent days of data consistently, with variability around self-correlation that matches that of real-life observed patterns for each data type and weekday type (weekday or weekend day). The space of possible scaling factors is discretised into $m$ intervals. The probability of transitioning from discrete factor intervals $i$ and $j$ is then:

$$p_{i,j} = \frac{n_{i,j}}{\sum_k n_{i,k}}$$

Where $n_{i,j}$ is the number of times that a transition between intervals $i$ and $j$ was recorded in subsequent days of data available.

As the probability of scaling factors is not evenly distributed between the minimum and maximum factors, we adopt a non-uniform discretisation approach based on percentile intervals, with finer data intervals for more common, lower scaling factors, and wider intervals for less common ones. This ensures that we retain granularity and information for more common lower factors. Furthermore, we use the 2D piecewise linear interpolation to fill in gaps in probability intervals, while ensuring the sum of probabilities for the next day always equals one.

Matrices of scaling factors transition probabilities $p_f(f_{t+1}|f_t, c_t, c_{t+1})$ are illustrated in Figure 9.

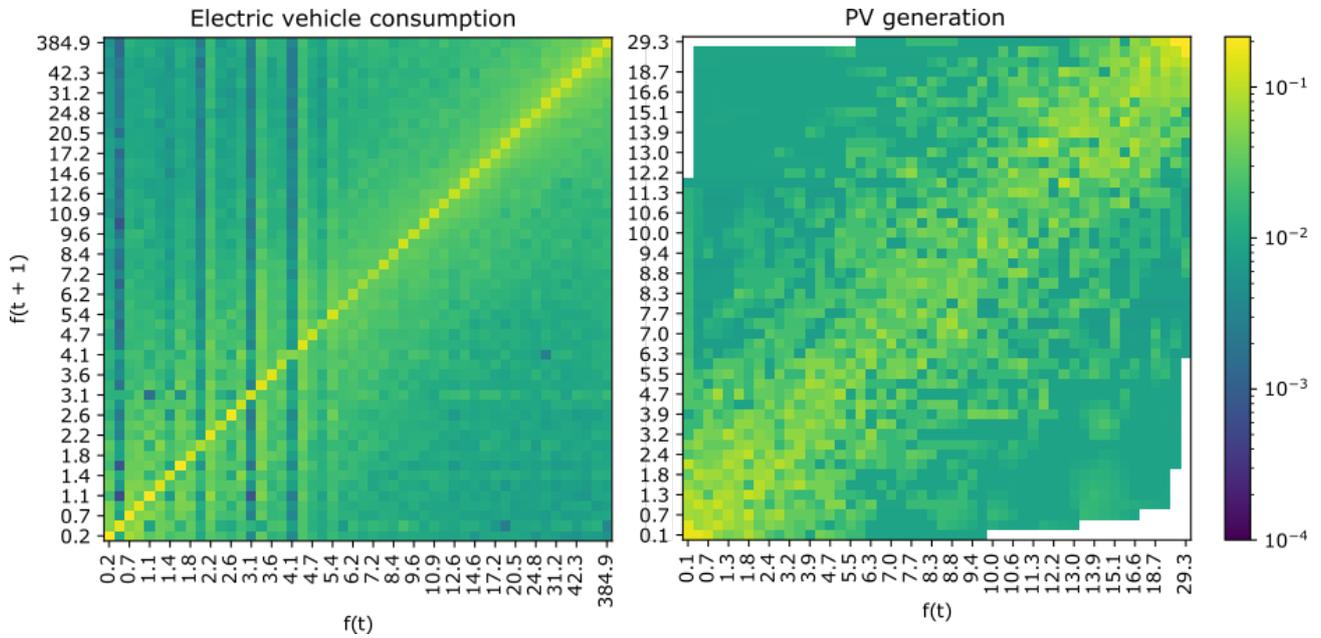

*Figure 9: Transition probability matrices between profile scaling factors in subsequent days (m=50 intervals).*

## 2.10. Behaviour cluster transition characterisation

In the case of behaviour-dependent data (household loads, EV patterns), we similarly characterise the probabilities $p_c(c_{t+1}|c_t, d_t, d_{t+1})$, of transitioning from one behaviour cluster to another in subsequent days for each day type transition ($d_t$ being weekday or weekend day), so that profiles can be generated using the adequately trained GAN networks. Variations in generated behaviour thus match real-life patterns for each new day.

## 3. Home Electricity Data Generator (HEDGE) tool

From the data processing described in Section 2, we obtain the following inputs for the Home Energy Data Generator (HEDGE):

a. Behaviour cluster transition matrices $P_c$
b. Normalised profiles generator (per data type, day type and behaviour cluster)
c. Scaling factors transition matrices $P_f$

Given an initial behaviour cluster and scaling factor for each home (which can be automatically selected in HEDGE to match their distribution in the original dataset), a Markov chain mechanism uses these to generate profiles for successive days, consistent across both scaling factors and behaviour clusters. The probabilistic Markov chain transition rules are:

1. For behaviour-dependent data types, select behaviour cluster $c_t$ based on the behaviour cluster transition matrix $p(c_{t+1}|c_t, d_t, d_{t+1})$, to select the appropriate GAN profile generator.
2. Generate a population of normalised profiles using pre-trained GAN weights, and randomly select one.
3. Scale the profile using a scaling factor according to the probabilities in the scaling factors transition matrices, from discrete distribution $p_f(f_{t+1}|f_t, c_t, c_{t+1})$

New random, realistic data can thus be generated for each subsequent day of simulation.

## 4. Limitations

A limitation of using this data generator to generate household data is that some heating and transport electrical loads may already be in the source data, as we do not have information on the breakdown of the household loads in the original dataset. There may therefore be a risk of double counting these loads if they are also modelled separately. However, the share of households using electric heating and possessing at-home EV chargers was low at the time of the data collection (2011-2014).

## Acknowledgments

This work was supported by the European Saven Scholarship, in partnership with the Department of Engineering Science, University of Oxford, and by UK Research and Innovation and the Engineering and Physical Sciences Research Council (award references EP/S000887/1, EP/S031901/1, and EP/T028564/1).